# DNA Hairpin Gate: A Renewable DNA Seesaw Motif Using Hairpins

*Abeer Eshra [1, 3*], Shalin Shah [1, 2] and John Reif [1, 2**]*
[1] Department of Computer Science, Duke University, Durham, NC, 27705, USA.
[2] Department of Electrical and Computer Engineering, Duke University, Durham, NC, 27705, USA
[3] Department of Computer Science and Engineering, Faculty of Electronic Eng., Menoufia University, Menouf, Menoufia, 32831, Egypt

## Abstract

In 2011, Qian et al. introduced the DNA seesaw gate motif, which is a powerful feed-forward DNA nanodevice that can perform digital logic computations. Their landmark work managed to evaluate moderately large Boolean circuits by cascading multiple DNA seesaw gates. Although their design is robust in solution and scalable, it is designed for one-time use and is not reusable. This prevents pursuing important applications such as feedback and sequential digital circuits. We present a novel design for DNA nanodevices that can perform digital logic computations and are furthermore renewable. First, we modified the prior DNA seesaw gate motif into a hairpin; we call the resulting motif a "DNA hairpin-seesaw gate". We show the feed-forward digital computation reaction imitates the seesaw gate motif. Second, we added a reporting phase that provides increased scalability to our device. Third, we designed input and fuel extracting hairpins that when added, initiate a renewing process. This results in a renewed functional gate, in its original configuration, which is able to make a new logical computation with new inputs. Finally, we introduced a renewable two-input Boolean logic OR gate. After calculating output of a certain input set, the circuit is restored and a new set of inputs is introduced to compute the new output. We provide experimental fluorescent data on three repeated rounds of executions of our hairpin gate motif and its restoration, indicating gradual loss of response. Finally, we calculated rate constants of our experimental data by fitting it to a second order reaction model using maximum likelihood estimation method.

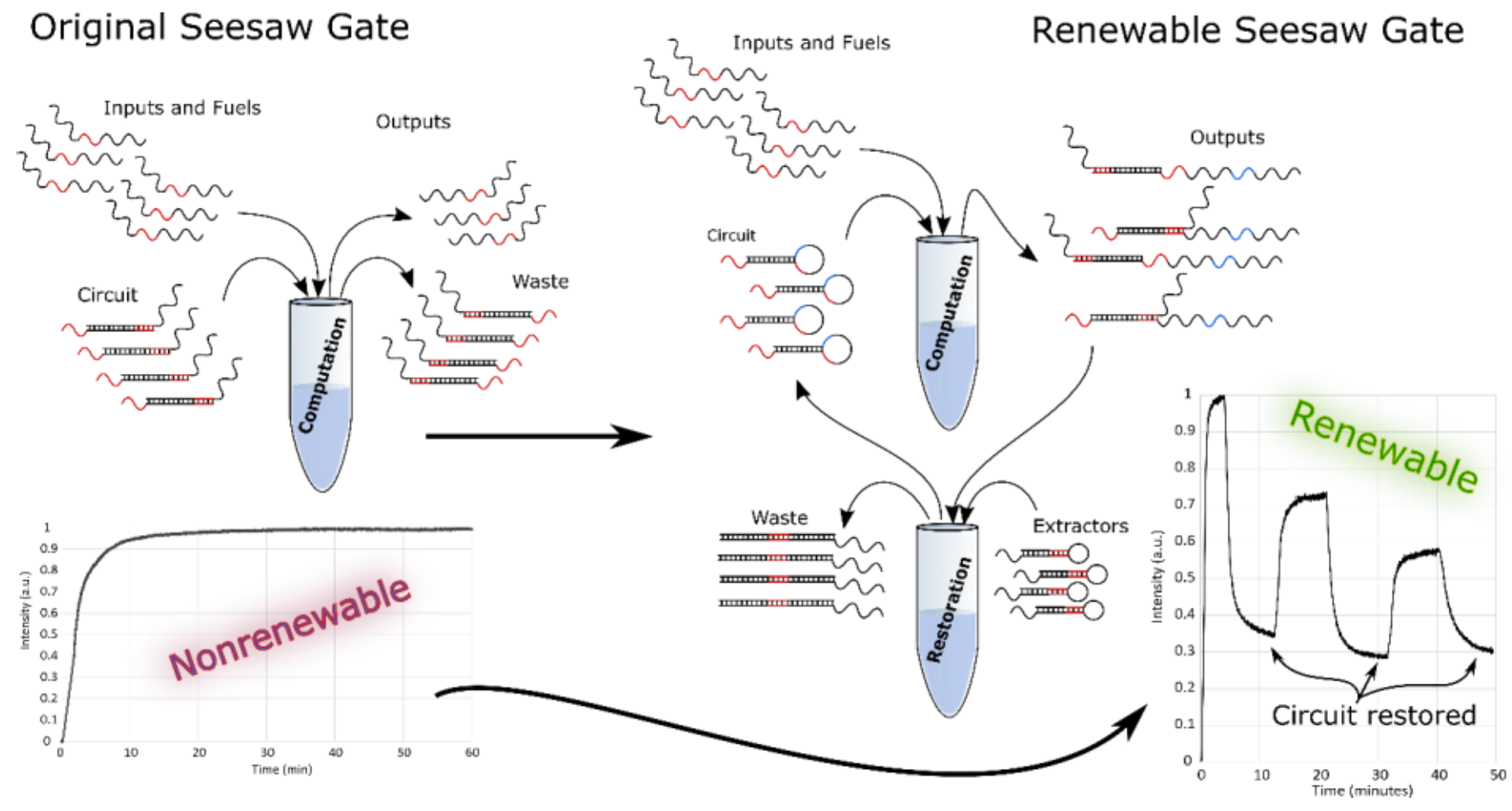

The field of DNA computing was initiated in 1994 by Leonard Adleman[1]. who first demonstrated the use of DNA as a computing machine by solving the Hamiltonian path problem. This was done by encoding the information in DNA sequences that were partially complement each another. Output was detected by observing bands on agarose gel electrophoresis. That experiment opened a door to many other applications and uses of DNA for computation. Researchers were able to demonstrate finite state machines[2-6], chemical reaction networks (CRNs)[7-9], Boolean logic circuits[10-15] and neural networks[16-17]. These DNA computations use DNA hybridization reactions, and in some cases, they made use of restriction enzymes[2, 18-19] or deoxyribozymes [6, 14, 20].

Most of these contributions only allowed one-time use of the DNA components in their computations. However, some prior works have demonstrated repeated use of DNA devices. For instance, Yurke *et al.* demonstrated a dynamic reusable DNA nanodevice that acted like a tweezer[21]. The movement of the DNA tweezer between its open and closed configurations was controlled by adding auxiliary strands that caused strand-displacement reactions: a fuel strand closed the tweezer, while its complement pulls the fuel and results in opening the tweezer again. This operation produces a double-stranded DNA waste that is accumulated at the end of every cycle. However, their device would help control motion to build nanostructures. In 2004, Sherman and Seeman developed a DNA nanodevice that acts like a bi-directional walker[22] which could walk in both directions. They self-assembled a one-dimensional DNA triple cross over (TX) tile nanostructure[23] to form a nano-track and self-assembled a two-legged structure that walked on the track. In the same year, Shin and Pierce also developed a bi-directional walker[24], which made use of fuel DNA strands to initiate the movement of each foot of their walker by strand-displacement reactions. In 2011, Goel and Ibrahimi implemented Boolean logic circuits based on restriction enzymes[25]. Their system was modular and renewable, but the use of restriction enzymes provided limitations, due to the relatively small number of available restriction enzymes and the waste accumulation due to the action of the restriction enzymes. In the same year, Chiniforooshan *et al.* proposed an enzyme-free theoretical design for DNA digital circuits[26]. Their design is scalable and time-responsive, energy-efficient, and renewable. However, to achieve all those advantages at the same time, their design was quite complex; along with the fan-out gates, they added a signal restoration gate to ensure input response on time for each gate. Also in 2011, Genot *et al.* developed a renewable three input logic circuit[27] that was non-catalytic and enzyme-free. However, reported experiments indicated relatively low fluorescence levels and slow execution time. Recently Garg *et al.* experimentally built renewable time responsive asynchronous logic circuits[28] that were also non-catalytic and enzyme-free. They used two designs to implement renewable gates, one using dsDNA complexes and the other using DNA hairpins. To achieve renewability, they used a pullback mechanism in which inputs are plucked to reconstruct the circuit. Later, new inputs are added to compute a new output. However, their reported experiments also indicated relatively slow execution time.

In the following, we let ssDNA denote a single-stranded DNA, let dsDNA denote a double-stranded DNA, and let a partial dsDNA be a DNA nanostructure resulting from the partial hybridization of two ssDNA. Note that a ssDNA can be used as a signal in a DNA computation, but when bound to its complement in a dsDNA, it is generally inactive. Toehold-mediated strand-displacement (developed by Yurke *et al.*[21]) is a reaction whereby a ssDNA, say S, invades a partial dsDNA, inserting S in place of a previously hybridized DNA strand. The toehold-mediated strand-displacement process begins with partial hybridization (see Figure 1.a) of the ssDNA S with the partially exposed part (called toehold) T of a dsDNA. Once the partial hybridization with T takes place, a branch migration reaction releases the originally bound strand. This method has been extensively studied theoretically[26, 29-30]. The reaction rate is controllable by the lengths of both toehold and branch[31]. DNA strand-displacement enabled DNA circuits that are enzyme free[21, 31-34], and instead the use of only DNA hybridization reactions. It was applied widely in practice, with many applications in molecular computations such as digital logic [35] and analog [36] circuits. On the other hand, toehold-exchange[31] (see Figure 1.b) is an alternative to toehold-mediated strand-displacement, which also solely involves hybridization reaction. Toehold-exchange makes use of two toeholds rather than one, where only one of the toeholds is exposed as single-stranded region at any time and the other toehold is sequestered within dsDNA. The toehold-exchange reaction results in a partial dsDNA with a new exposed toehold on the product DNA. Toehold-exchange has two potential advantages over toehold-mediated strand-displacement. (i) The first advantage is that the resulting new single-stranded region allows further toehold-exchange reactions, in contrast to toehold mediated strand-displacement (Note that toehold mediated strand-displacement results in a fully double-stranded DNA which can be expected to be inactive with respect to further hybridization reactions.). (ii) The second advantage is both the forward and backward reactions are generally faster.

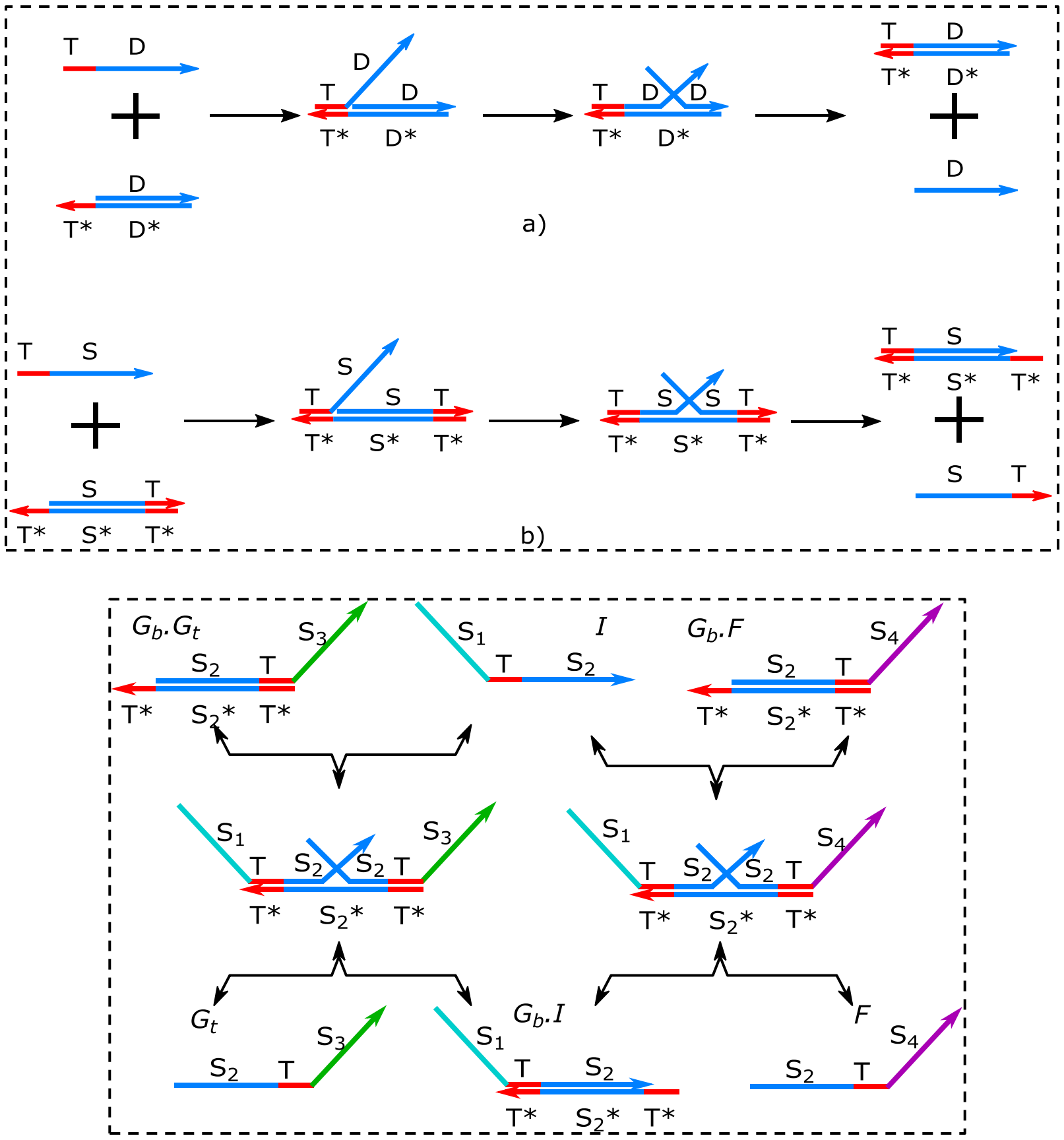

c)

**Figure 1:** Strand-displacement mechanisms. (a) Toehold mediated strand-displacement. (b) Toehold-exchange strand-displacement. (c) Basic seesaw reaction. Input $I$ invades the gate to release output. Fuel $F$ replaces input $I$ to work as a catalyst and release more output. This is a complete catalytic cycle. (c) was reproduced with permission from Qian *et al.* [37]

In 2011, Qian *et al* presented a novel DNA gate motif [37] based on toehold-exchange mechanism. They called it seesaw gate because the reaction keeps going forward and backward until it reaches equilibrium. Two seesawing steps complete a catalytic reaction. Figure 1.c shows the reaction of a seesaw gate motif. With toehold-exchange strand-displacement, input signal $I$ invades the gate complex $G$ by a toehold-mediated strand replacement reaction, to release output $O$. The fuel $F$ works as a catalyst, which begins replacing input strand $I$ from the former reaction resulted complex. This helps input invade other gates to release more outputs. Once the reaction reaches equilibrium, the output is calculated by the resulting concentration. Two predetermined range of concentrations are interpreted as logical values (logical 0 and logical 1, respectively). This allows the input and output of the seesaw gate to be interpreted as logical values. Threshold complex is used to restore the concentration of the outputs to these concentration ranges.

Qian's seesaw gate motif is a powerful feed-forward DNA nanodevice that resulted in robust and scalable circuits. There are two reasons for such characteristics: (i) First, they used a universal toehold for all strand-displacement cascades, allowing easy transition from one seesaw motif to another. (ii) Second, their system is catalytic, which aids in the amplification of small signals through the circuit, and increases reaction rates. It was used to design an AND gate and an OR gate, and furthermore a dual-rail design[38] was used to represent any circuit using only AND and OR gates. Experimentally demonstrated circuits using seesaw gates included a four-bit square root circuit[13] and a four-neuron Hopfield associated memory. Although it is a powerful motif, it still lacks a renewability feature. Once the reactions of a circuit of seesaw gates reach steady state and the result is collected, everything in the solution in now non-reusable (*i.e.* waste).

In this work, we report a novel design for a renewable DNA logical gate. The seesaw motif is modified to be a hairpin and is called a hairpin-seesaw gate motif. This strategy packs the two reaction cycles in one gate motif and helps regeneration of the gate once the reversal process starts. Two toeholds instead of one were utilized. In addition, a reporting phase was added to increase scalability. The proposed system was demonstrated to be reusable multiple times. Fluorescence spectrometer and native gel imaging were used to report the experimental results.

# Results and discussion

*Our Renewable Hairpin-Seesaw Motif:* DNA hairpins with long stems are extremely stable especially when there are no mismatches within the hairpin double-stranded stem[39]. The original two-strand DNA complex seesaw motif was modified to a hairpin that is renewable. Therefore, the gate's stem is the double-stranded region of the seesaw motif. Its loop is the single-stranded right arm that cascades to the next stage. In addition, two universal toeholds as opposed to one type were used. As seen in Figure 2.a, $G$ is the gate, $I$ is the input, and $F$ is the fuel. $G$ is a hairpin where the left toehold $T_1$ interacts with its complement in the middle of $I$ to open the hairpin. As in the original seesaw motif, the reaction starts with $I$ displacing the top strand of the hairpin stem with regular strand-displacement. This results in a partial dsDNA complex $G.I$. When the hairpin is open, the middle toehold $T_1$ becomes vulnerable to hybridization. The fuel strand $F$ initiates a strand exchange and replaces $I$, resulting in complex $G.F$. Therefore, the fuel $F$ aids $I$ to work as a catalyst, invade new gates and produce more output. However, the use of two universal toeholds can potentially increase leaks, because one of the two toeholds in the strand exchange is supposed to be sequestered. In our design, having $T_1^*$ single-stranded inside the loop can allow $F$ to invade the gate before being opened by $I$. This potential leak problem could be solved by making the two universal toeholds partially complementary to each other, so to disallow complete hybridization between the toeholds and to partially sequester the inner toehold as well. Hence, the fuel strand is incapable of invading the gate before input. Figure 2.a shows the basic forward reaction discussed above.

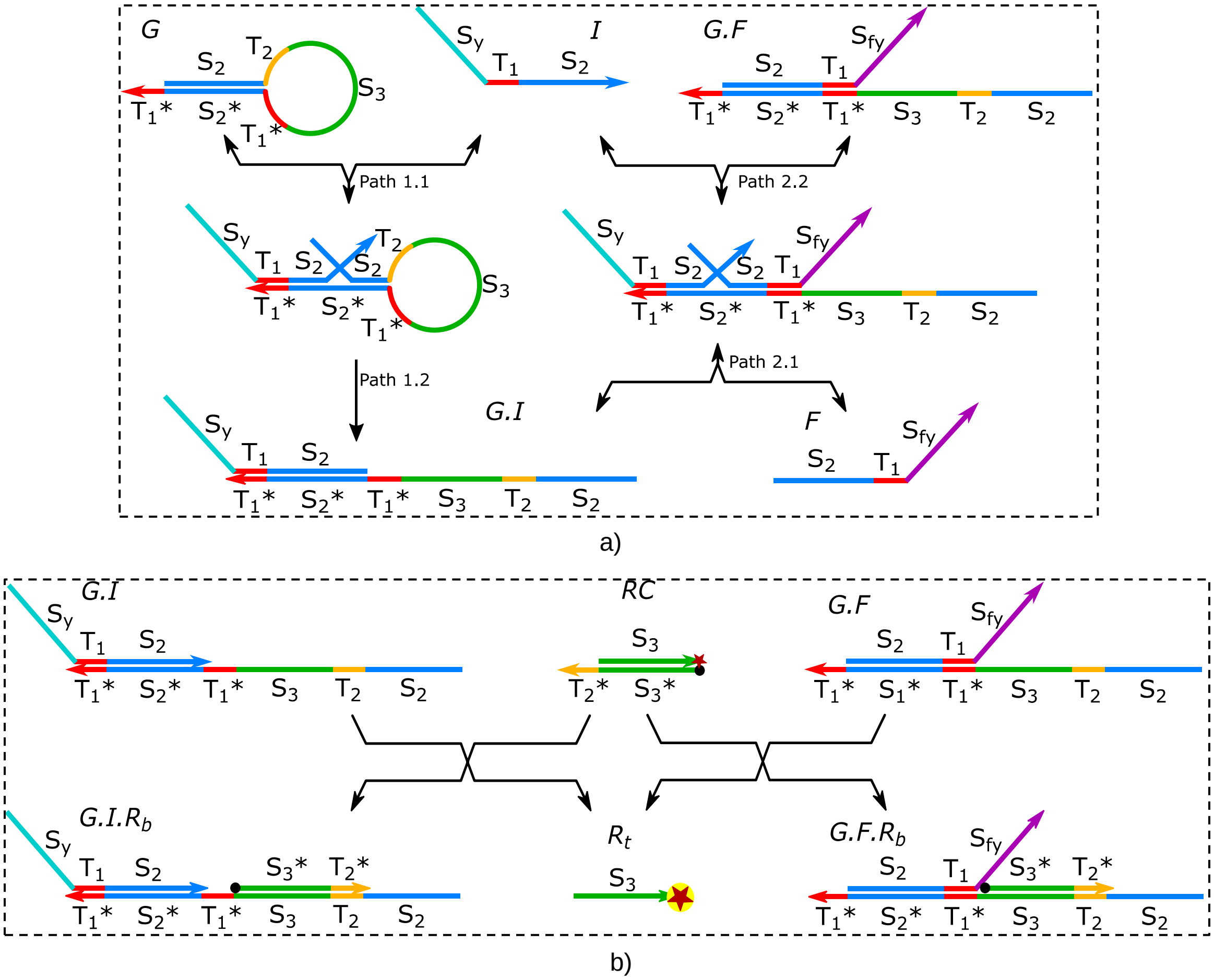

**Figure 2:** (a) Basic hairpin-seesaw catalytic cycle. Input binds to gate (G) then is replaced by fuel to work as a catalyst and binds to other gates. Path 1: Input binds to the gate to open the hairpin. Path 2: Fuel binds to the input-gate complex to release input and form fuel-gate complex. Arrows indicate the reaction pathways either forward or reverse. (b) Reporting mechanism. Another cascade was added to improve system scalability. Reporter complex is a dsDNA with a toehold, which has a complement within the hairpin loop. Once hairpin is opened by either input or fuel, toehold $T_2$ is available to react. Via toehold-mediate strand-displacement, the reporter complex hybridizes to the gate. As a result, the dye and quencher molecules are separated, causing the dye to fluoresce. This fluorescence emission indicates the reaction completion. The reporter complex hybridizes to the gate-input or gate-fuel complexes. The fluorescence emission is an indication of the completion of the hybridization.

To detect the output and give the system more scalability, another reaction stage was added after the gate reaction, via a reporter complex $R$. It is modified with a pair of fluorophore and quencher molecules. The reporter $R$ is a duplex with a toehold $T_2$ that is complementary to the third toehold in the hairpin. As seen in Figure 2.b, once the gate is open the third toehold is exposed, and $R$ begins interacting with either $G.I$ or $G.F$ by toehold-mediated strand-displacement. This process releases the final detectable output.

*Renewing the Hairpin Seesaw Motif:* To renew the hairpin-seesaw gate, we initiate the process of reversal of the forward reaction by adding extracting hairpins. They have a complementary toehold to start strand-displacement with either input or fuel. These extractors remove both input and fuel strands, freeing the hairpin stem regions from hybridization. The resulting strand replacement reaction on the hairpin-seesaw gate starts from the ends of the hairpin stem arms until reaching a point that weakens the prior hybridization of the attached reporter strand. At this point, a subsequence of the reporter strand becomes single-stranded, which in turn acts as an active toehold to initiate strand-displacement. This reaction will reconstruct both the hairpin-seesaw gate and the reporter complex.

Since both input and fuel have their own distinctive arms, we used strand-displacement as an extraction mechanism. As seen in Figure 3, two extracting hairpins were introduced. Each extractor can attach to either the input strand $I$ or fuel strand $F$. For brevity, we will discuss the extraction of the input strand $I$ (the same scenario would take place with the fuel strand $F$) *using* extractor $I_{ex}$. The effect of extractor $I_{ex}$ is that two strand-displacement processes take place consecutively:

(i) The first happens when the extractor $I_{ex}$ start hybridizing with $I$ free arm, and $I$ starts opening the extractor hairpin by displacing stem top strand of $I_{ex}$.

(ii) The second strand-displacement starts when $I_{ex}$ hairpin is open, it starts displacing $G$ from being hybridized with $I$.

The entire extraction process ends by $I$ being extracted, *e.g.* via conventional magnetic bead extraction[40]. A similar process is used to extract the fuel strand $F$.

After extracting both the input and fuel strands, the hairpin motif have two free arms that contains two domains complementary to each other. The restoration process can be expected to proceed as follows:

(i) The hybridization process of the hairpin stem arms starts from the ends.

(ii) When approaching the loop (which is hybridized to $R$ bottom strand) edges of the hybridized part will experience breathing effect and open toeholds from $R$. This is due to the strength of stem hybridization.

(iii) This will allow the reporter top strand to start hybridizing with its complement pulling it away from the hairpin loop.

(iv) Finally, this will reform the hairpin and restore the reporter complex.

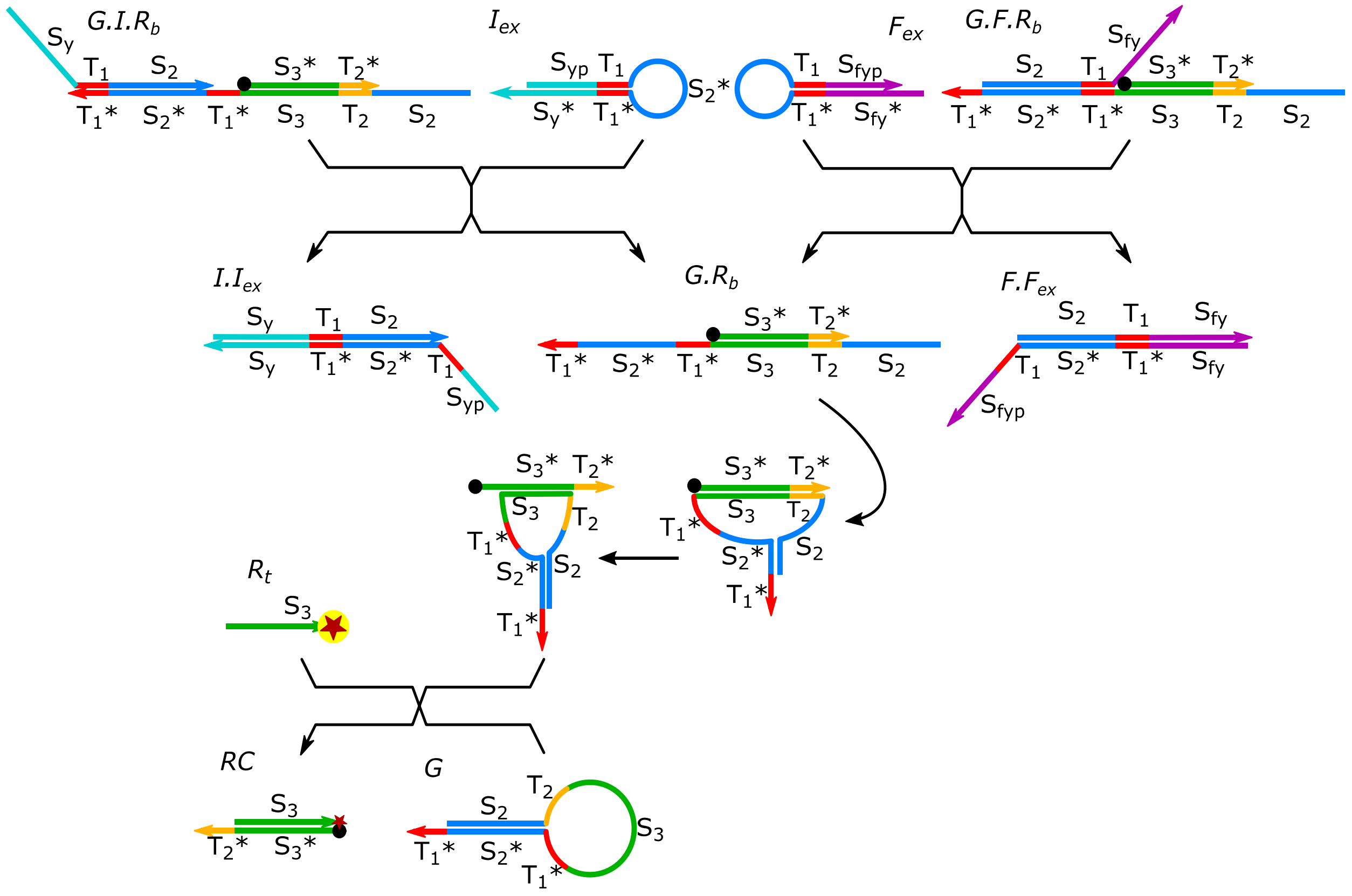

**Figure 3:** Reverse Mechanism. Using extraction mechanism, two extractors ($I_{ex}$ and $F_{ex}$) were introduced to pull input and fuel out from forward reaction resultant complexes. Once extracted, gate hairpin tries to close up starting from the ends of its branches. This weakens hybridization with the reporter bottom, and opens a toehold for the reporter top to hybridize with its complement by strand-displacement. Extracted input and fuel become waste. Reporter and gate are fully restored to be reused. Detailed extraction steps are shown in supplementary Figure S1.

The hairpin-seesaw gate motif was experimentally verified to be renewable. We used fluorescence spectroscopy to test our system. The reporter was modified with a fluorophore-quencher pair. Figure 4 shows results of renewing the motif three times. Both kinetic experiment and polyacrylamide gel electrophoresis analysis show the gradual loss in signal. Those experiments were carried out in three phases. In the following $1x$ equals $100\ nM$. In first phase, $G$ and $R$ were mixed in solution with concentrations of $1x$ and $1.5x$ respectively. $I$ and $F$ were then added to solution with concentrations of $1x$ and $2x$ respectively, allowing the forward reaction to take place. When signal reached peak saturation, we initiated the reversal process by adding $2x$ of $I_{ex}$ and $2x$ of $F_{ex}$. After having output its lowest saturation level, the second phase was initiated by adding $I$ and $F$ in double concentration of $I_{ex}$and $F_{ex}$ to reverse the backward reaction and to initiate the forward reaction again. The same process was repeated with doubling concentrations of inserts every time. Experiments were performed in Tris-acetate-EDTA buffer with 12.5 $mM$ $Mg^{+2}$ at 22 $°C$.

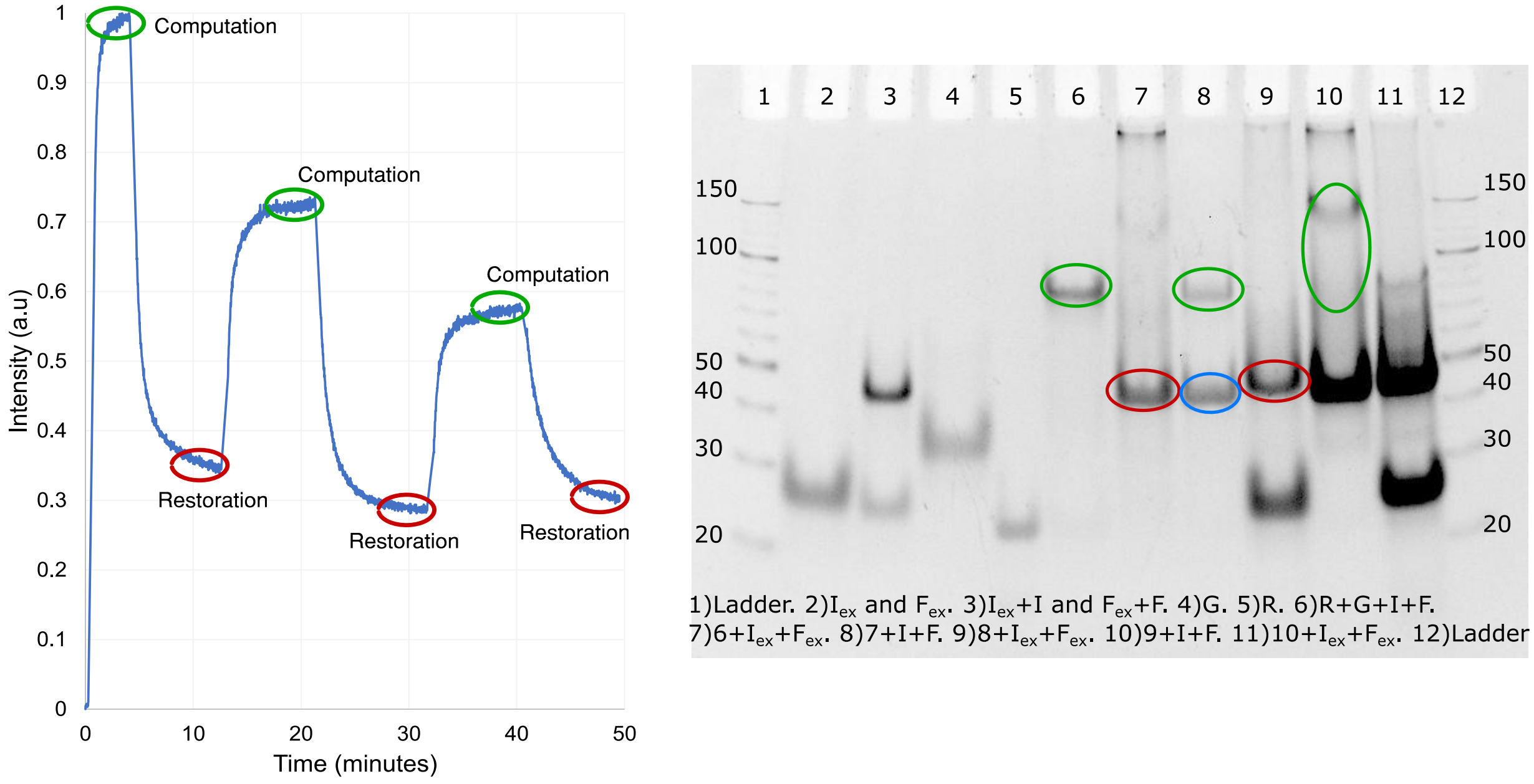

a) b)

**Figure 4:** System recycled three times with gradual loss in signal. (a) Kinetic experiment. (b) Polyacrylamide gel electrophoresis analysis. All visible bands are DNA with double-stranded regions. In both figures, red circles show result of adding extractors to perform restoration. Green circles show result of adding Input and fuel, which start the forward reaction and reuse the circuit. In (b) a blue circle is showing accumulating waste from the previous restoration. Intensity of adjacent band in lane 10 shows that concentration of waste is increasing.

*Our Renewable OR Gate:* Using the presented reversible motif, we designed a two-input OR gate illustrated in Figure 5.a and b. It consists of two hairpin gates that act in parallel. If one of the inputs is present, one of the gates will be opened to work with the reporter and release the output. In case both the inputs are present, both gates should open and output will be released. In case there is no input, none of the gates will open which means there should be no output release. The four cases were experimentally verified and renewed in Figure 5.c. To prove that it is reusable, we computed a combination of inputs, generated the output, restored the gate and then recomputed with a different combination of inputs. Two different experiments are shown in Figure 5.d and e.

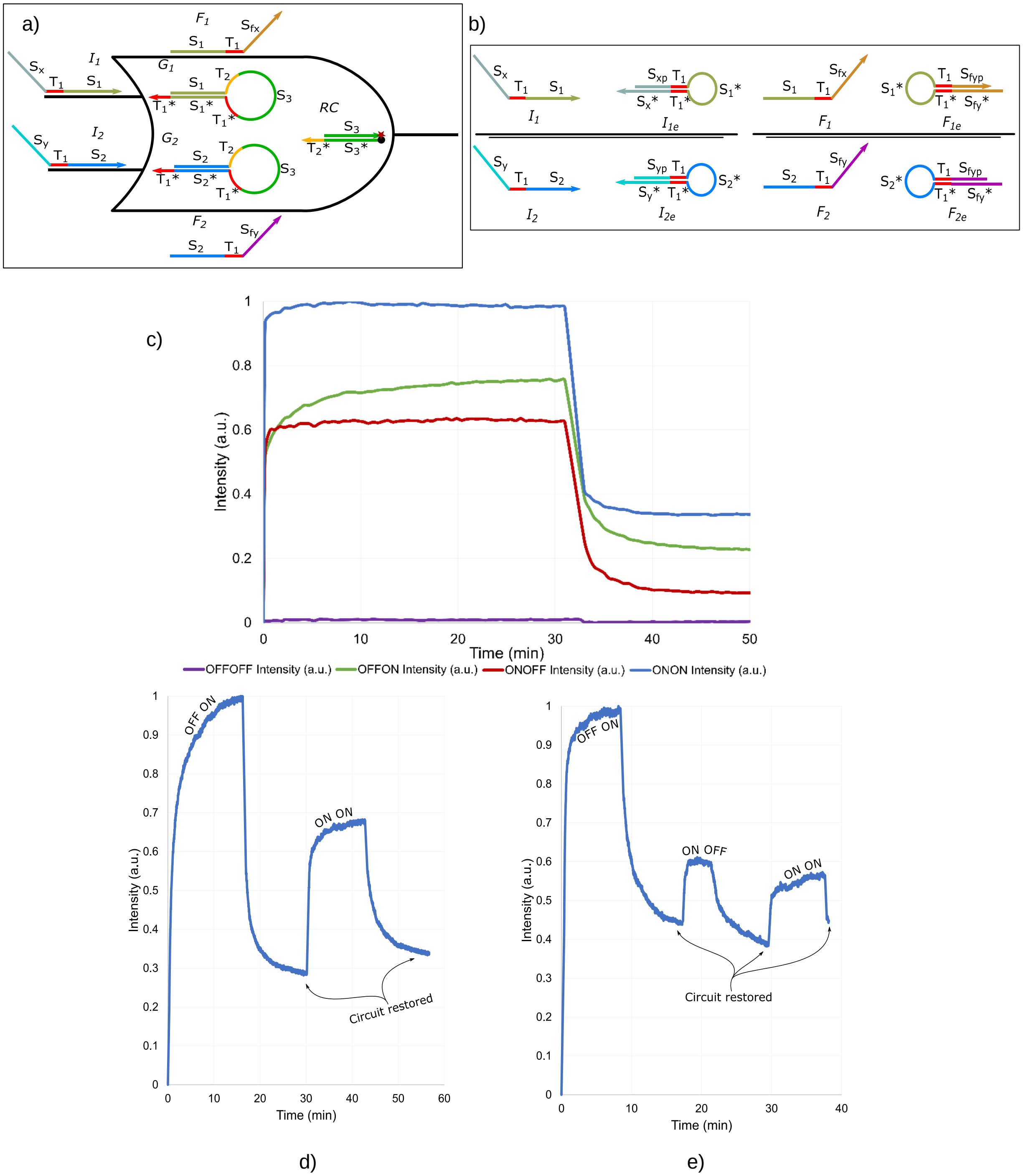

**Figure 5:** Reversible 2-input OR gate. (a) Abstract design. (b) Inputs and fuels each in-lined with their extractor. (c) Four cases computation and restoration. OR gate should give high output if one of its inputs is high. It should give low output if all inputs are low. Fluorescence was absent only when there was no inputs. Time is given for pipetting. (d and e) OR gate with reversal and changing cases. (d) OFFON then ONON. (e) OFFON, ONOFF then ONON.

*Modeling of the System's Reaction Kinetics:* To understand our system better and identify potential leaks, we modelled the reaction system by assuming each strand is a single molecule, which can only bind to desired complementary strand or undergo toehold mediated strand-displacement. It is assumed that the reporter rate constant $k_{rep} = 1.3\ x\ 10^6/M/s$, as reported in previous works [31, 41]. Additionally, we also assume that the rate constant for all other reactions is the same and we denote it as $k_t$. We model our reversible motif with a set of reactions provide in supplementary section S3.

We used Microsoft's Language for Synthetic Biology (LBS)[42] package to model and simulate our set of reaction systems. To obtain the value of rate constant $k_t$, we used maximum likelihood estimation (MLE) with initial values reported in Qian *et al.*[13] and the range of $k_t$, over which optimization problem was solved, was specified as $10^5$: $10^6$ /M /s as reported by Zhang *et al.*[31]. MLE is common statistical technique where a model is defined and likelihood of model-parameters is calculated over a range of values to find a maximum value. The calculate value of rate constant value for best-fit of the model is $2.743\ X\ 10^6/M/s$. Note that we assumed full yield of products for first cycle and changed available concentration of input, fuel, input extractors and fuel extractors in the subsequent cycles to achieve a good fit between our model and data. This loss in reaction yield has reported in previous studies [21, 43]. The yield of our system drops after every cycle, as observed from gel and electrophoresis data, and therefore our model needs to account for that drop. Reasons for loss includes poisoned gates after each cycle, either computation or restoration. Also, in each cycle, adding more DNA liquid to solution causes dilution of previously existed DNA. This affects the output signal as well. A solution to that is rescaling and considering the highest obtained signal is our maximum with every cycle. This is not addressed within this work.

Additionally, we used the same model to fit data for OR gate and obtained a rate constant of $2.45\ X\ 10^6/M/s$ which is relatively close to the rate constant for motif. LBS fitting code for the basic motif three cycle reversal is provided in supplementary section S5. Figure 6 shows experimental and simulated data of basic motif three cycles of reversal and OR gate four cases computation and restoration.

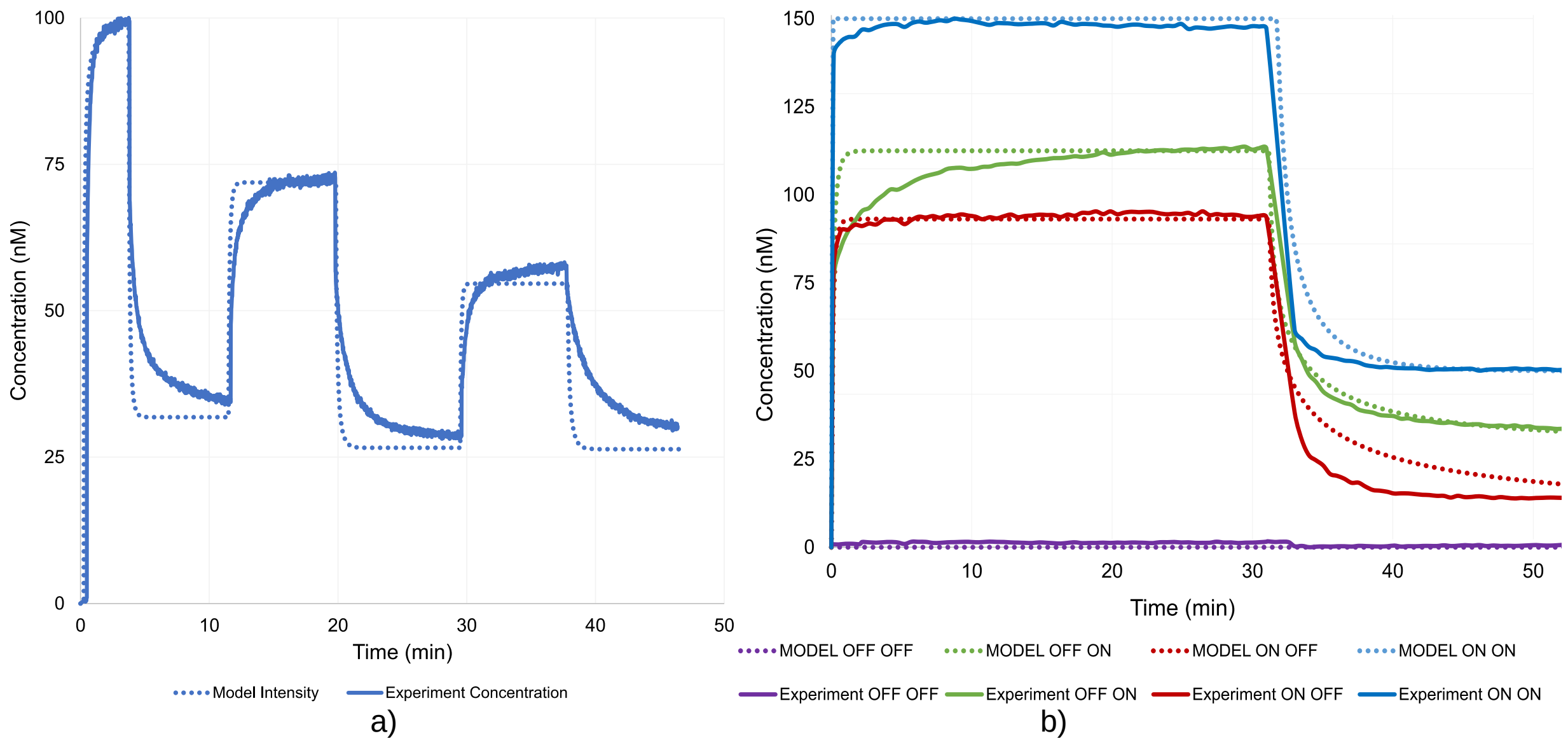

**Figure 6:** Experiment modeling. (a) Modeling of three restoration cycles of the motif. Rate constant obtained is $2.743\text{x}10^{6}/\text{M}/s$. (b) Modeling of OR gate computation and restoration of four cases. Rate constant obtained is $2.45\text{x}10^{6}/\text{M}/s$. In both figures, solid lines are experimental results and dotted lines are modeling results.

*System Design:* Since the hairpin loop contains a domain $S_i$ and two toeholds ($T_i$ and $T_{i+1}$), there was a chance of cross-talk with other ssDNA. To minimize the cross-talk effect, we used 3-letter encoding (A, C, and T) for every ssDNA. G in particular was avoided in ssDNA. Guanine base (G) has two forms enol form and keto form. The dominant one is the keto form when G binds naturally with C. However, when G is in enol form it binds to T[44-45]. Using 3-letter (ACT) coding ensures that any G base is confined within a double-stranded region and can only interact when the double helix is invaded by strand-displacement.

To limit the interaction between the fuel and gate before input invasion, we intended making a two base complementarity between $T_1{}^*$ and $T_2$ as illustrated in Figure 7. The two toeholds are 5 nt long. Two base-pairs are not strong enough to hinder the forward reaction from opening[46]. However, they could impede the fuel strand from interacting with the hairpin before opening.

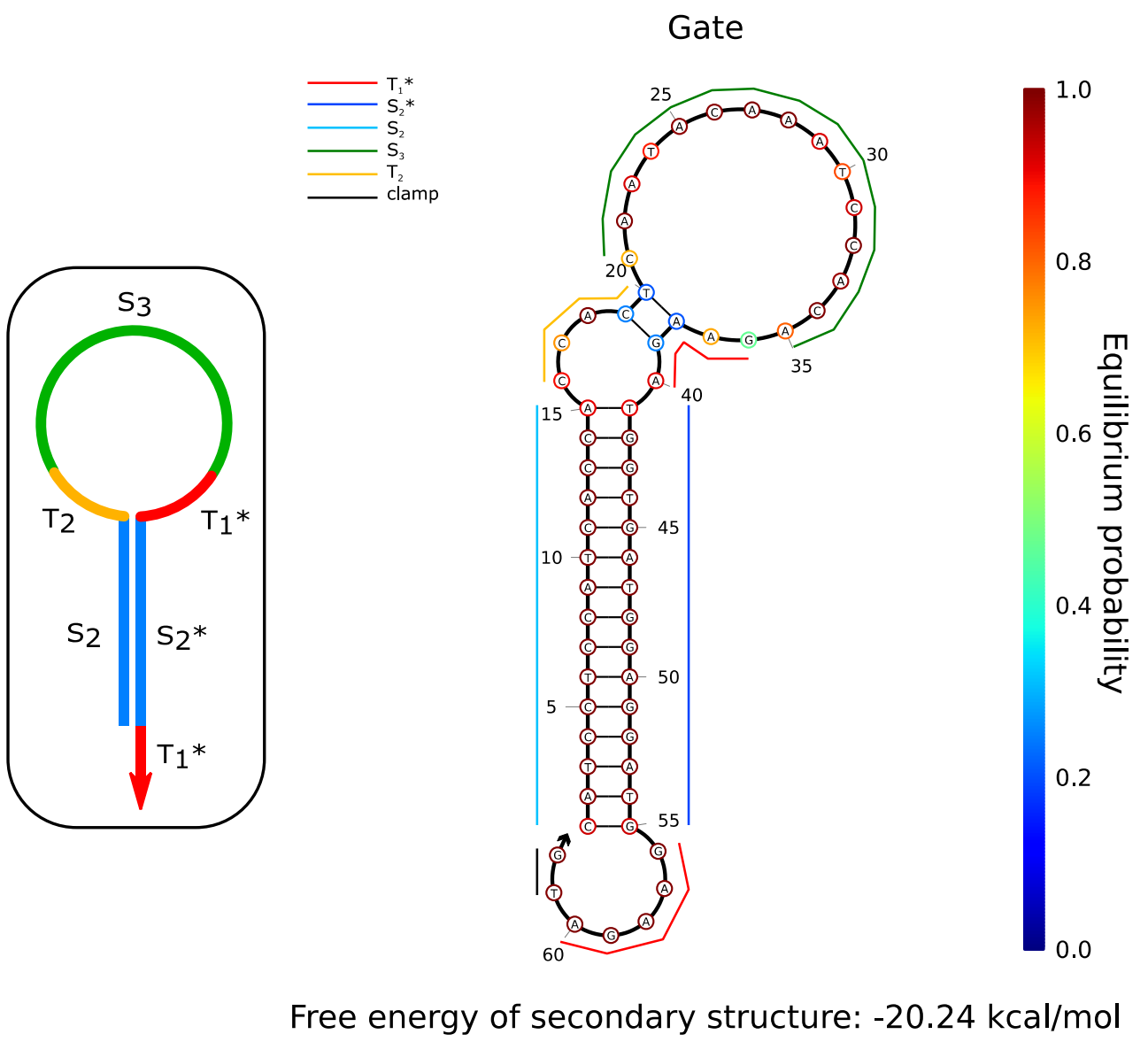

**Figure 7:** Basic gate design shown by abstract and detailed representation

Domain sequences were adopted from Qian *et al.*[13] and manually modified to give stable structures. NuPack [47] was used to verify the desired gate structure and correct hybridization between intended domains. We also studied the fuel effect on the reaction with the hindering site. In supplementary, Figure S2 shows that the fuel boosts the output level because of the catalytic reaction. However, there is some leak due to the partial sequestering.

*Conclusion:* In this work, we have addressed the problem of designing renewable enzyme-free DNA devices. Our main goal was to change DNA computing from use-once to be renewable. We managed to make the basic seesaw motif renewable. To achieve this: (i) we modified the seesaw motif into a hairpin and used extractors to pull out every gate invader, (ii) we added a separate reporting stage that has two roles: (a) reporting the output and (b) providing scalability of the system, (iii) both gate and reporter are restored after the reversing process and shall be reused by adding new input and fuel.

To demonstrate the general applicability of our idea we introduced a 2-input OR gate. We presented the four cases of computation and restoration. We also used the same circuit to compute different cases consecutively. With adding time control, our system can be time-responsive as well. The designs were experimentally validated with both reaction kinetics by observing fluorescence, and polyacrylamide gel imaging by analyzing bands under UV light. Experimental results showed that renewability is possible with our circuits. Our next goal is to scale our design by demonstrating large DNA circuits. In the future, we intend to use this DNA gate restoration technique to build enzyme-free DNA sequential circuits that take inputs and outputs from cells, to be used in cell programming, medical diagnoses and therapeutics.

## Material and Methods

All unmodified sequences were ordered unpurified from IDT Integrated DNA Technologies. Only the modified strands of the reporter were ordered HPLC purified and delivered as $100\ \mu L$ with $100\ \mu M$ concentration. All other strands were delivered dehydrated. Sequences are designed as shown in Table 1.

**Table 1:** DNA Sequences of basic reversible seesaw motif and reusable OR gate. (Note that toehold $T_1$ is colored Blue, toehold $T_2$ is colored Red, Extractors' toeholds are highlighted grey, the two complementary bases between $T_1$ and $T_2$ within the hairpin are highlighted in yellow

| Circuit | strand | Sequence |
|---|---|---|
| ***Basic reversible seesaw motif*** | $F$ | CATCCTCCATCACCA TCTTC CATTTTTTTTTTTCA |
| | $I$ | CAACACCTACATCCA TCTTC CATCCTCCATCACCA |
| | $G$ | CATCCTCCATCACCA CCACT CAATACAAATCCACA GAAGA TGGTGATGGAGGATG GAAGA TG |
| | $F_{ex}$ | TGAAAAAAAAAAATG GAAGA TGGTGATGGAGGATG TCTTC CATTTTTTTT |
| | $I_{ex}$ | CCTACATCCA TCTTC TGGTGATGGAGGATG GAAGA TGGATGTAGGTGTTG |
| ***OR gate*** | $F_1$ | CAACTCATAATTCCA TCTTC CATTTTTTTTTATAC |
| | $I_1$ | CAACACCTACATCCA TCTTC CAACTCATAATTCCA |
| | $G_1$ | CAACTCATAATTCCA CCACT CAATACAAATCCACA GAAGA TGGAATTATGAGTTG GAAGA TG |
| | $F_2$ | CATCCTCCATCACCA TCTTC CATTTTTTTTTTTCA |
| | $I_2$ | CACAATCACACACCA TCTTC CATCCTCCATCACCA |
| | $G_2$ | CATCCTCCATCACCA CCACT CAATACAAATCCACA GAAGA TGGTGATGGAGGATG GAAGA TG |
| | $F_{1ex}$ | GTATAAAAAAAAATG GAAGA TGGAATTATGAGTTG TCTTC CATTTTTTTT |
| | $I_{1ex}$ | CCTACATCCA TCTTC TGGAATTATGAGTTG GAAGA TGGATGTAGGTGTTG |
| | $F_{2ex}$ | TGAAAAAAAAAAATG GAAGA TGGTGATGGAGGATG TCTTC CATTTTTTTT |
| | $I_{2ex}$ | TCACACACCA TCTTC TGGTGATGGAGGATG GAAGA TGGTGTGTGATTGTG |
| ***Reporter complex*** | $R_t$ | CAA TAC AAA TCC ACA CCG /3IABkFQ/ |
| | $R_b$ | /56-FAM/ CGGTGTGGATTTGTATTG AGTGG |

60 $\mu L$ of nuclease free water and 60 $\mu L$ of $2x$ denaturing dye were added to dilute the powder. Solutions were given a water bath on 95 ℃ for 10 minutes then left to cool down in room temperature. Afterwards strands were purified using 10% Denaturing PAGE for 90 min on 300 V using $1x$ TBE as a buffer. When gel finished running, DNA bands were excerpted under UV and soaked in elution buffer overnight. Further steps included a butanol wash, incubation in 200 proof ethanol in −80 ℃, 70% ethanol wash and drying the liquids on 30 ℃ for 2 hours using a vacuum centrifuge method to get purified DNA in powder form.

After ssDNA purification, hairpins were formed using a PCR thermocycler. They were heated up to 95 ℃ then slowly cooled down to 20 ℃. Reporter complex consists of two strands. Top strand was modified with a quencher (Iowa Black® FQ from IDTDNA) on the 3' end and bottom strand was modified with a dye (6-FAM (Fluorescein) from IDTDNA) on the 5' end. Within our illustrating figures (Figure 2, Figure 3 and Figure 5) was reversed intentionally to be dye on the top strand and quencher on the bottom strand. This was done to simplify demonstrating output collection. Reporter also was conjugated with the same PCR procedure. Both reporter strands were mixed with $1.2x$ excess of the top strand. There are two reasons for this choice: (i) First is that the top strand is modified with quencher which will not influence the fluorescence measurement if there is excess of it. (ii) Second the toehold to achieve strand-displacement is included within the bottom strand, insuring that every reporter strand with toehold is well covered in order not to interfere with the gate loop except after being open by strand-displacement. Concentrations were measured using NanoDrop, ND-1000. Stock solutions were prepared for every species with three levels of concentrations, low, medium and high.

*Kinetic Experiment:* Output fluorescence intensity of kinetic experiments were measured using Cary eclipse Varian spectrophotometer (Duke University, Durham, NC). Temperature was set on 22 °C for all experiments. Excitation and emission wavelengths were 495 and 520 $nm$ respectively. Measure amount of each on desired concentration in total volume of 80 $\mu L$ (using this equation: $conc_1 * v_1 = conc_2 * v_2$). TAE $Mg^{+2}$ was used as a reaction buffer. Initial signal was obtained after mixing $R$ and $G$. Later, samples were pipetted rapidly in the cuvette to obtain a stable fluorescence. Order of pipetting and required volumes are shown in supplementary Table S1.

*Native Gel Electrophoresis:* Regarding the gel in Figure 4.b, lanes 1 and 12 contain 6 $\mu L$ of ready-to-use 20 bp DNA Ladder, from ThermoFisher. All other lanes contain samples diluted in TAE $Mg^{+2}$ with a yield of 1 $\mu M$ concentrations in 50 $\mu L$. Lane 2 contains both $I_{ex}$ and $F_{ex}$. Lane 3 was prepared in two Eppendorf tubes separately. One contained $I$ and $I_{ex}$ and the second contained $F$ and $F_{ex}$. Both were incubated in room temperature for 1 hour. Afterwards, they were mixed together in one tube. Lane 4 contains $G$ solution. Lane 5 contains $R$ solution. Lane 6 through lane 11, seven samples that were prepared as identical solutions containing $R$ and $G$. All were kept at room temperature. Afterwards, $I$ and $F$ were added to all tubes and incubated in room temperature to react for 20 minutes. Starting from Lane 7 $I_{ex}$ and $F_{ex}$ were added in double concentrations of previously added $I$ and $F$. Again, all tubes were incubated at room temperature for 20 minutes. Same process was repeated with $I$ and $F$ or $I_{ex}$ and $F_{ex}$ with doubling concentrations on every insertion and leaving a tube out. Later a 10% native polyacrylamide gel was and casted. $6x$ native dye was added to all samples and loaded into gel wells. Gel was run at 150 $V$ for 5.5 hours in $1x$ TBE buffer. Afterwards, it was stained for 30 minutes in solution containing 0.5 $\mu g/mL$ of ethidium bromide and de-stained for another 30 minutes in deionized water. Gel was imaged under UV light to show DNA bands.

## Associated Contents

**Supporting Information Available.** It includes Table S1, Figures S1-S2, simulation code and kinetic reaction equation (.pdf). This material is available free of charge *via* the Internet at http://pubs.acs.org.

## Author Information

**Corresponding authors:**

*abeer.eshra@duke.edu;

**reif@cs.duke.edu

**Notes:**

Authors declare no competing financial interest.

## Funding Sources

This work was supported by Missions & Cultural Representation Sector, Ministry of Higher Education, Egypt and also by NSF CCF1617791.

## ACKNOWLEDGEMENT

We thank Dr. Hieu Bui for his insightful and intuitive discussions and revisions. A. Eshra sincerely thanks Prof. Nawal El-Fishawy and Ass. Prof. Ayman El-Sayed for their support and cooperation.